\documentclass[twocolumn,prb,citeautoscript,superscriptaddress,a4paper,floatfix]{revtex4-1}
\usepackage{graphicx}
\usepackage{amssymb}
\usepackage{amsmath}
\newcommand{\etal}{\emph{et al.}}
\newcommand{\ie}{\emph{i.e.}}

\newcommand{\sto}{SrTiO$_{3}$}
\newcommand{\TO}{Ta$_{2}$O$_{5}$}

\newcommand{\pary}{Parylene-C}
\newcommand{\Lo}{\mathop{L_{\mbox{\scriptsize o}}}}						
\newcommand{\ID}{\mathop{I_{\mbox{\tiny SD}}}}							
\newcommand{\VD}{\mathop{V_{\mbox{\tiny SD}}}}						
\newcommand{\VG}{\mathop{V_{\mbox{\tiny G}}}}							
\newcommand{\Vth}{\mathop{V_{\mbox{\scriptsize th}}}}					
\newcommand{\DV}{\mathop{\Delta V}}									
\newcommand{\muFE}{\mathop{\mu_{\mbox{\tiny FE}}}}					
\newcommand{\ssq}{\mathop{\sigma_{\mbox{\scriptsize sq}}}}				
\newcommand{\nsq}{\mathop{n_{\mbox{\scriptsize sq}}}}					
\newcommand{\csq}{\mathop{C_{\mbox{\scriptsize sq}}}}					
\newcommand{\df}[2]{\mathop{\displaystyle\frac{#1}{#2}}}
\newcommand{\psidis}{\mathop{\psi_{\mbox{\scriptsize dis}}}}				
\begin{document}
%
\title{%
Enhanced and continuous electrostatic carrier doping on the \sto\ surface.
}
%
%
\author{A. B. Eyvazov}
\affiliation{Division of Physics and Applied Physics,
Nanyang Technological University, 637371, Singapore.}
\affiliation{National Institute of Advanced Industrial Science and Technology (AIST),
Tsukuba 305-8562, Japan.}
\author{I. H. Inoue$^{*}$}
\affiliation{National Institute of Advanced Industrial Science and Technology (AIST),
Tsukuba 305-8562, Japan.}
\affiliation{CREST, Japan Science and Technology Agency (JST), Tokyo 102-0075, Japan.}
\author{P. Stoliar}
\affiliation{CNRS - Laboratoire de Physique des Solides, Universite Paris-Sud, Orsay 91405, France.}
\affiliation{ECyT, Universidad Nacional de San Mart\'in, San Mart\'in 1650, Argentina.}
\author{M. J. Rozenberg}
\affiliation{CNRS - Laboratoire de Physique des Solides, Universite Paris-Sud, Orsay 91405, France.}
\author{C. Panagopoulos}
\affiliation{Division of Physics and Applied Physics,
Nanyang Technological University, 637371, Singapore.}
\date{\normalsize 28th December 2012, Revised 15th March 2013}
%
\begin{abstract}
Paraelectrical tuning of a charge carrier density as high as 10$^{13}$\,cm$^{-2}$ in the presence of a high electronic carrier mobility on the delicate surfaces of correlated oxides, is a key to the technological breakthrough of a field effect transistor (FET) utilising the metal-nonmetal transition.
Here we introduce the \pary/\TO\ hybrid gate insulator and fabricate FET devices on single-crystalline \sto,
which has been regarded as a bedrock material for oxide electronics.
The gate insulator accumulates up to $\sim$10$^{13}$cm$^{-2}$ carriers,
while the field-effect mobility is kept at 10\,cm$^2$/Vs even at room temperature.
Further to the exceptional performance of our devices, the enhanced compatibility of high carrier density and high mobility revealed the mechanism for the long standing puzzle of the distribution of electrostatically doped carriers on the surface of \sto.
Namely, the formation and continuous evolution of field domains and current filaments.
\end{abstract}
\maketitle
%
%
\section*{Introduction}
A prominent feature of phase transitions in correlated electron systems is the nanoscale dynamics leading to future electronics such as switching devices and novel nonvolatile resistance-change memory applications \cite{chakhalian,hwang,hormoz,inoue1,xiang,asanuma,nakano,zyang,yang}.
In spite of the long-standing research, device prototypes remain limited mainly due to the absence of a method for a continuous (\ie, paraelectrical) and precise electrostatic control (gating) of a two dimensional (2D) charge carrier density $\gtrsim$10$^{13}$\,cm$^{-2}$.
Furthermore, oxygen and cation stoichiometry are fairly unstable in perovskite-type (ABO$_{3}$) transition-metal oxides (TMOs) \cite{phillips}.
Deposition of dielectric oxides on the surface of ABO$_{3}$ channel for gating inevitably results in such kind of defects, introducing randomness and eventually smearing the critical features of metal-insulator transitions \cite{miyashita}. 
In the face of these challenges, we have chosen to fabricate FET adopting single-crystalline \sto\ for the channel. 
%
\begin{figure}[!hb]
\includegraphics[width=\columnwidth,clip]{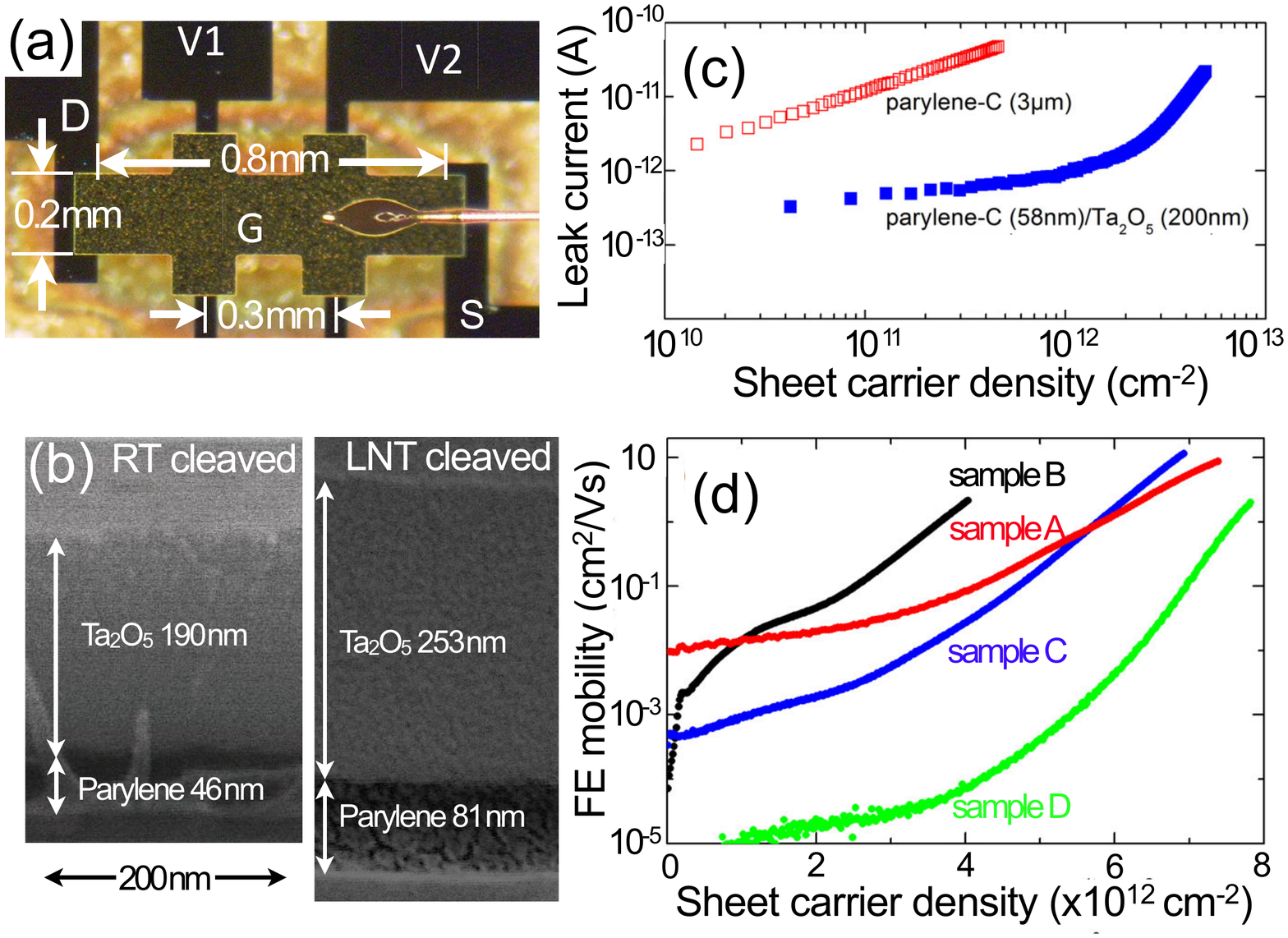}
\caption{
\label{images} (a) Photograph of a typical FET structure fabricated on the top of a (100) surface of a single-crystalline \sto.
Gate (G), source (S), drain (D) electrodes and four other potential probes such as V$_{1}$ and V$_{2}$ are indicated.
A gold wire of 25\,$\mu$m diameter is attached by conducting gold paint (Silvest 8560-1A, Tokuriki Chemicals).
(b) Scanning electron microscopy images of the cross sections of the bilayer gate insulator.
The sample was cleaved at room temperature (left) and at liquid N$_{2}$ temperature (right).
See Methods Summary for details.
(c) Gate leak current plotted against the sheet carrier density for the bilayer gate insulator with 200\,nm \TO\ deposited on the top of 48\,nm \pary\ (blue opaque square) and for the single layer gate insulator of 3\,$\mu$m \pary\ (red transparent square).
(d) Field-effect (FE) mobility plotted against the sheet carrier density for samples A, B, C, and D with different \pary\ thickness as summarised in Table~\ref{table1}.
}
\end{figure}
This compound is a band insulator with a crystal structure characteristic of many TMO Mott insulators, and becomes metallic by electron doping.
\sto\ is an n-type wide-gap semiconductor with a band gap of 3.2eV.
It is considered as a matrix for spintronic devices due to the unique configuration of the spin-precession vector at the surface and the absence of the Dresselhaus term \cite{nakamura2012}.
The surface of the single crystal can be atomically flat and is widely used as a substrate for correlated TMO thin films, making it preferable candidate to test the suitability of the reported gate insulator for future electronics.
Identifying the technical methods and physical mechanism for paraelectric tuning of electronic conduction in two dimensional electron gases (2DEG) created at the surface of \sto\ is a problem of bursting interdisciplinary interest \cite{dagotto}.

High-density charge accumulation $\sim$\,$10^{15}$\,cm$^{-2}$ was recently realised \cite{yuan} through an electric double-layer (EDL) on the surface of electrolytes such as ionic liquids.
Although the gating with EDL is a promising method for future correlated electron devices, its viability is hindered, because ionic liquids on the surface of TMOs may trigger redox reactions, causing vacancies at the interface especially in the presence of high electric fields \cite{ueno2010-1,ueno2010-2}.
In addition, the continuous control of the carrier density with EDL can be done only by using liquid electrolytes, which are unsuitable for integration to the present solid-state electronics.

\begin{table}
\caption{
\label{table1} \pary\ thickness of the four different samples reported in this work. The values of thickness were estimated from the measured values of capacitance of the bilayer gate insulators.
}
	\begin{tabular}{cccc}
	\hline\hline
	\parbox{0.15\columnwidth}{\strut{}Sample\strut}%
	 & \parbox{0.25\columnwidth}{\strut{}Amount of raw \\ material\,(mg)\strut}%
	 & \parbox{0.25\columnwidth}{\strut{}Bilayer capacitance \\ per area (nF/cm$^2$)\strut}%
	 & \parbox{0.25\columnwidth}{\strut{}Estimated \pary\ thickness\,(nm)\strut}\\
	\hline
	A & 300 & 34 & 57.7\\
	B & 350 & 24.97 & 87.9\\
	C & 400 & 22.25 & 101.7\\
	D & 600 & 12.56 & 199.9\\
	\hline\hline
	\end{tabular}
\end{table}

Here, we have engineered a bilayer of poly-monochloro-\emph{para}-xylylene (known as \pary) \cite{gorham} with \TO, Fig.~\ref{images}.
\TO\ has a high dielectric constant $\varepsilon$\,=\,25 and a relatively high dielectric strength $\sim$4\,MV/cm \cite{sethi}.
The combination of \TO\ and \pary\ into a single gate insulator enabled a sheet carrier density of $\sim$10$^{13}$\,cm$^{-2}$ while keeping the field effect mobility of the \sto\ surface at $\sim$10\,cm$^{2}$/Vs.
Fabrication details can be found in the Methods Summary.
The electronic transport measurements were performed using an Agilent 4155C semiconductor parameter analyser.

\section*{Results}
Fig.~\ref{images}\,(c) shows the gate leak currents plotted against the sheet carrier density for the bilayer gate insulator and for the single \pary\ gate insulator.
Although the total thickness of the bilayer gate insulator is 10 times thinner than the 3\,$\mu$m-thick \pary, the leak current of the bilayer is an order of magnitude lower, which is similar to the trend reported earlier \cite{deman}.
Hence, the sputtering deposition of \TO\ does not damage the \pary\ thin film.
Fig.~\ref{images}\,(d) depicts an upturn in the increase of the field effect mobility at sheet carrier densities $2 - 4\times 10^{12}$\,cm$^{-2}$.
The field effect mobility is defined as $\muFE\equiv\df{1}{e}\df{d\ssq}{d\nsq}$, where $\ssq\equiv(\ID/W)/(\DV\!\!/\Lo)$ is the sheet conductance and $\nsq\equiv\csq(\VG-\Vth)/e$ is the nominal sheet carrier density in the channel region.
$\ID$ is the channel current, $\Lo$ (and $\DV$) is the distance (and voltage difference) between channel electrodes V$_{2}$ and V$_{1}$, $W$ is the channel width, $\csq$ is the measured total capacitance per unit area, and $e$ is the elementary charge.
Although $\ID$ increased almost exponentially with the application of $\VG$, the $\ID$-$\VG$ behaviour is different from that in the conventional subthreshold regime. 
In fact, all our samples showed extremely large $\muFE$, $\sim$10\,cm$^2$/Vs at large carrier densities (\ie, large $\VG$) without reaching saturation even at room temperature.
Thus, we set the threshold voltage $\Vth$\,=\,0.
Notably, $\muFE$ is an order of magnitude larger than earlier reports for a \pary/\sto\ FET at room temperature ($\muFE$\,=\,0.3\,cm$^2$/Vs for $\nsq$\,=\,3$\times$10$^{12}$\,cm$^{-2}$) \cite{nakamura3}, and comparable to the bulk Hall mobility of \sto\ \cite{ahrens}.

\begin{figure}[!hb]
\includegraphics[width=\columnwidth,clip]{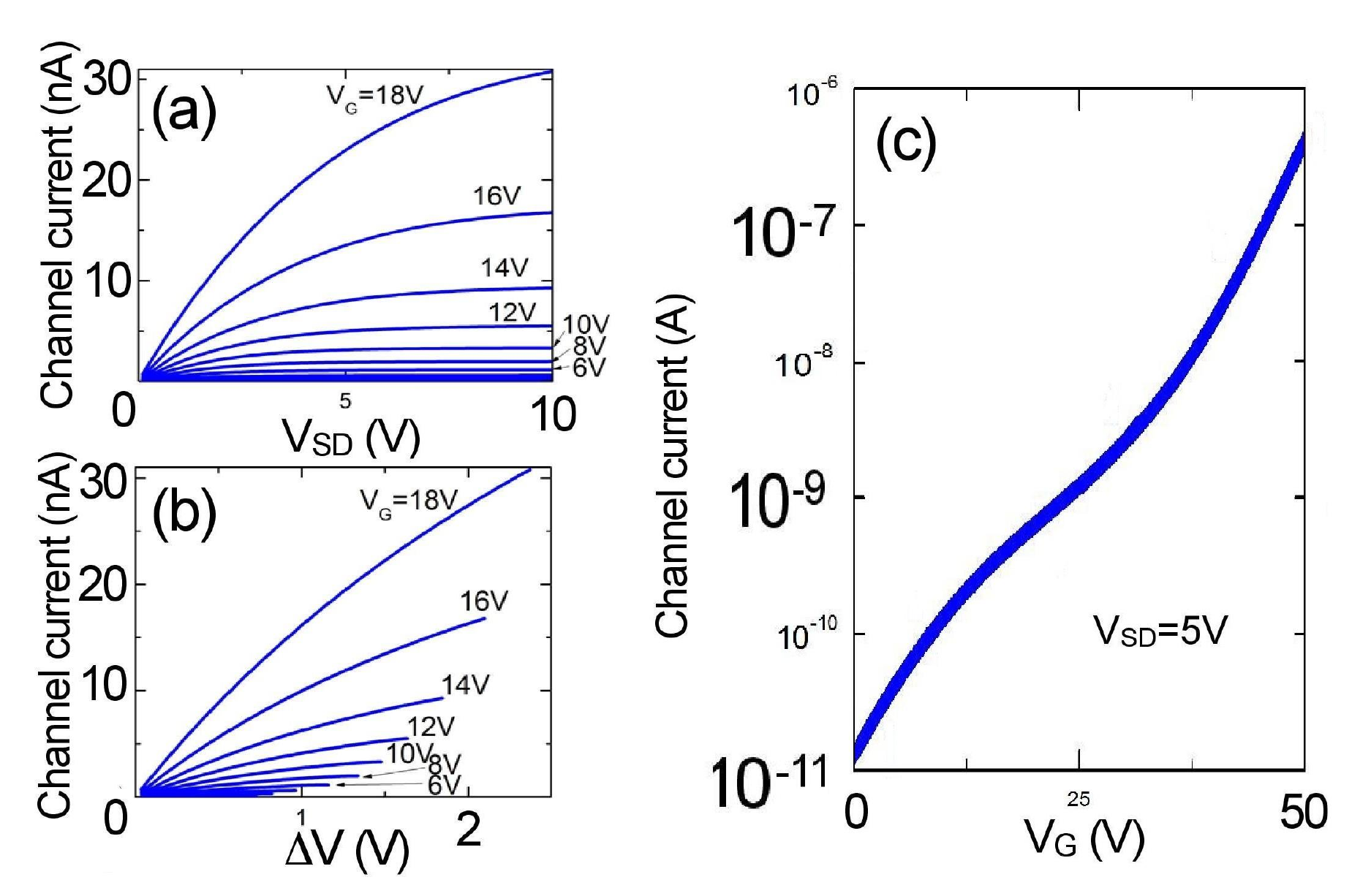}
\caption{%
\label{FETcharact}
(a) Channel current of Sample\,B plotted against the source-drain voltage $\VD$ for the different values of the applied gate voltage $\VG$.
(b) Same as (a) but the horizontal axis is the voltage difference $\DV$ between V$_{2}$ and V$_{1}$ electrodes.
(c) Channel current of Sample\,C plotted against $\VG$ for $\VD$\,=\,5\,V.
}
\end{figure}

The typical $\ID$-$\VD$ characteristics of Sample B in Fig.~\ref{FETcharact}\,(a) indicate that the current saturates for large $\VD$ as in conventional semiconductor FETs, where it occurs due to the pinch-off of the channel.
The saturation vanishes when the channel current in Fig.~\ref{FETcharact}\,(a) is plotted against the voltage drop inside the channel $\DV$, Fig.~\ref{FETcharact}\,(b).
Hence, the pinch-off region is rather small even at low $\VG$ and emerges between the drain and the V$_{2}$ electrode for $\VD$ below 5\,V.
Fig.~\ref{FETcharact}\,(c) shows that the current increases monotonically from $\sim$1\,nA to $\sim$100\,$\mu$A as the gate voltage increases up to 40\,V, thereby demonstrating a continuous electrostatic control of the sheet carrier density.

\begin{figure}[!htb]
\includegraphics[width=\columnwidth,clip]{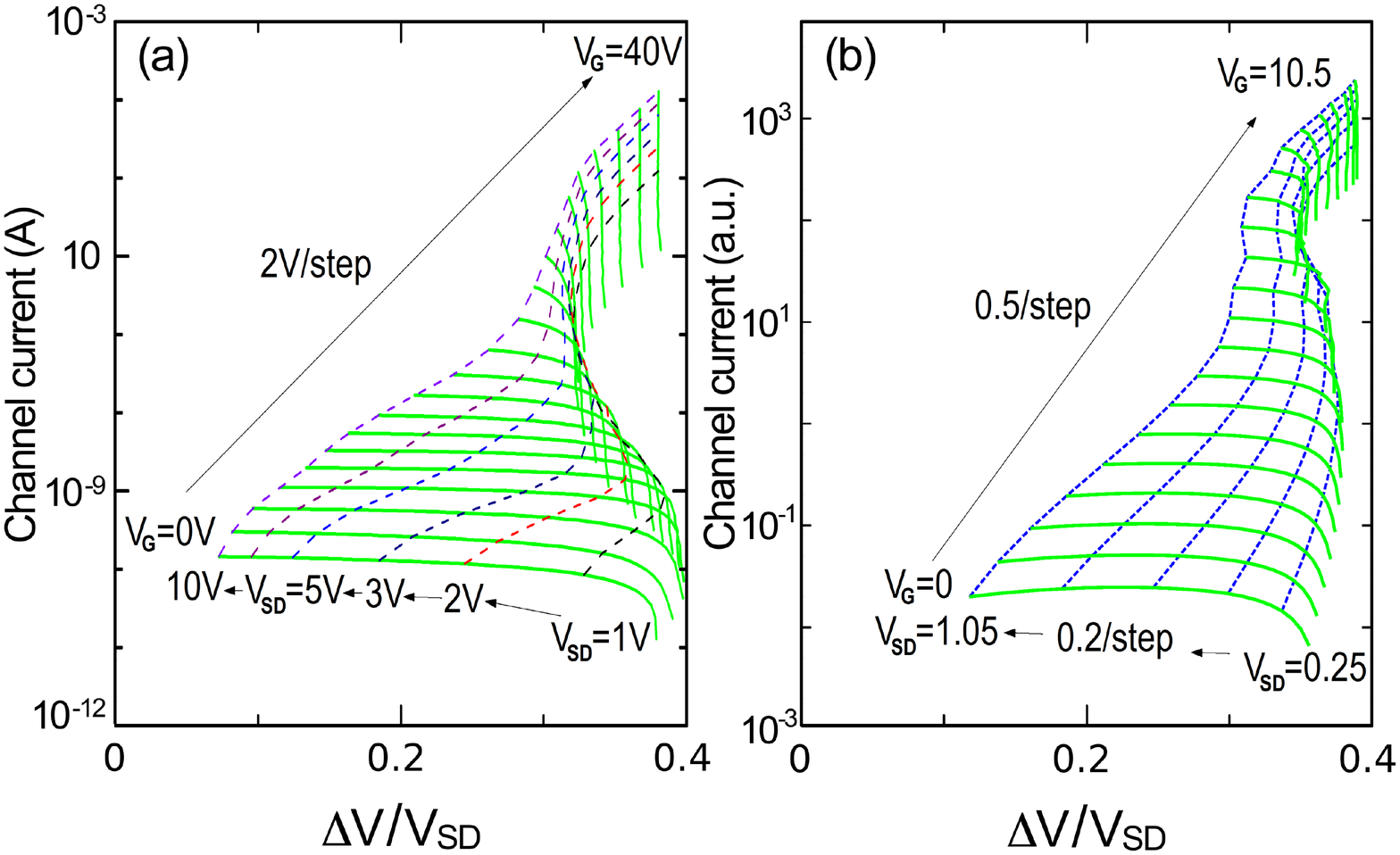}
\caption{%
\label{S-shape}
(a) Channel current of Sample C plotted against $\DV$/$\VD$.
$\VD$ is swept from 0 to 10\,V for the different values of $\VG$ as shown in the thick green lines.
The dashed lines indicate the $\VD$ contour, in which SNDR is seen for low $\VD$ and vanishes with increasing $\VD$.
(b) The results of numerical simulations. The voltages are expressed in arbitrary units. See Supplementary Materials for details.
}
\end{figure}

Furthermore, for small fixed values of $\VG$, although the channel current does not change very much (Fig.\,\ref{S-shape}\,(a)), the ratio $\DV/\VD$ decreases rapidly with increasing $\VD$.
This feature can be associated with the formation of field domains \cite{ridley}.
We also observe that in various parameter conditions, $\DV/\VD$ approaches the 
geometrical value $\sim$0.37\,=\,(V$_{1}$-V$_{2}$ distance 295$\mu$m)/(source-drain distance 800\,$\mu$m).
This indicates that in those regimes the contact resistance between \sto\ and Al electrodes, which is expected to be ohmic, is indeed negligibly small compared to the channel resistance of \sto. 
It is interesting, however, to observe the behaviour of $\DV/\VD$ with increasing $\VG$ and with fixed (small) value of $\VD$ in Fig.\,\ref{S-shape}\,(a).
This feature is the so-called S-type negative differential resistivity (SNDR) associated with the formation of current filaments \cite{ridley}. 
%

\begin{figure}[!htb]
\includegraphics[width=\columnwidth,clip]{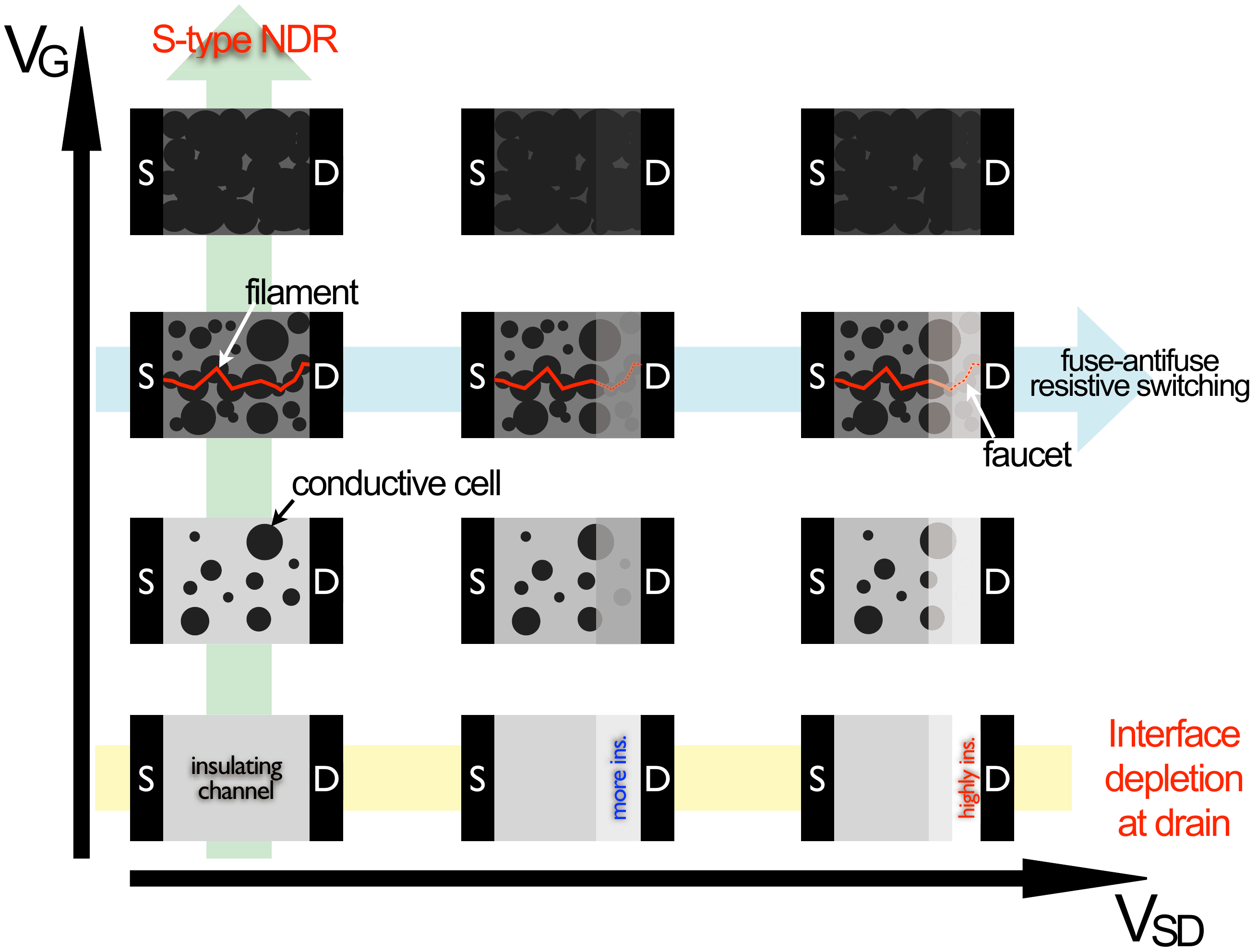}
\caption{%
\label{inhomo}
Schematic pictures of the evolution of conductive cells in the channel as demonstrated in Fig.\,\ref{S-shape}.
For increasing $\VG$, the channel varies from insulator to relatively good metal as the size of the conductive cells increases and a percolating current filament is formed.
This represents SNDR.
Meanwhile, for increasing $\VD$, the region near the drain electrode is pinched off.
By controlling the two parameters $\VD$ and $\VG$, we can realise the formation and continuous evolution of field domains and current filaments; \ie, the competition of SNDR and the interface depletion at drain, which is not only the main topic of the research for the resistance change memory \cite{yang} but also a common problem associated to the future oxide electronics.
The filament can be ruptured by the Joule-heating at the interface to the metal electrode (called faucet \cite{inoue2}); one of the widely accepted models of the non-volatile resistance change (fuse-antifuse resistive switching \cite{yang}).
}
\end{figure}

The rather complicated behaviour of the $\ID$ vs $(\DV\!/\VD)$ curves in Fig.\,\ref{S-shape}\,(a) was qualitatively reproduced (Fig.\,\ref{S-shape}\,(b)) by a model calculation based on the following assumptions: (i) the existence of a low $\VG$ regime where charge carriers are mainly spatially and inhomogeneously confined in cells, and move through the channel region by incoherent tunnelling, (ii) a higher $\VG$ regime where the conductive cells overlap and percolating conduction paths are formed, and (iii) a well developed homogeneous 2DEG with barriers formed at the interfaces.
We modelled the experimental system through a 2D resistor network array where the transport properties of each cell is determined primarily by the local electron density.
The cells are assumed to be randomly distributed shallow potential wells for carriers with a site dependent potential strength $\psidis({\bf r})$.
At low $\VG$ and $\VD$, the cells have low density and the carriers are confined. 
The carriers move between cells by a variable range hopping (VRH) mechanism.
At higher applied voltages, the occupation of cells increases and the carriers delocalise, augmenting their mobility.
For given external potentials $\VG$ and $\VD$, we solve for the resistor network and compute the local voltages at all cells along the 2D channel.
These local voltages are then compared to the corresponding values of the local confining potential of the cell $\psidis({\bf r})$.
If their difference is negative, the mobility is assumed to be the VRH type, whereas if it is positive, we adopt a much larger value, consistent with a band-like conduction in the metallic state.
The interfaces at the source and drain electrodes are simply modelled as barriers with potential height $\psi_b$.
The details of the model and the numerical study are presented in the Supplementary Materials.
Our numerical study captures the qualitative transport behaviour shown in Fig.\,\ref{S-shape}, providing new insight to the conduction states of 2DEGs in \sto. 
In fact, it demonstrates that the system evolves through a variety of regimes connected through crossovers depicted schematically in Fig.\,\ref{inhomo}.

\section*{Discussion}
In summary, we have demonstrated that the \pary/\TO\ bilayer overcomes previous limitations, achieving the required large electrostatic carrier doping, while keeping a high quality interface with the TMO.
The hybrid gate insulator can sustain a sheet carrier density of 8$\times$10$^{12}$\,cm$^{-2}$, without introducing formidable damage to the channel interface, as the field-effect mobility of \sto\ exceeds 10\,cm$^2$/Vs, even at room temperature.
The excellent agreement of our numerical study with measurements on the hybrid gate insulator, indicate a continuous evolution through a variety of different regimes, including tuneability of field domains and current filaments on the surface of \sto.
Our findings provide a novel method for paraelectrical tuning of metal-nonmetal transitions in correlated electron devices made of ABO$_3$-type TMOs. 
Notably, the fabrication details and physical mechanism reported here may be applied towards emergent oxide electronics controlled by low dimensional charge carrier transport. 
%

Further to the practical interest, low temperature studies on similar devices would reveal whether granular superconductivity \cite{beloborodov} in \sto\ could be realised for small $\VG$ and $\VD$.
Indeed, the behaviour of possible superconducting cells with dimensions comparable to characteristic length scales may be considerably different from bulk materials.
Moreover, the quantum confinement in the cells may lift the degeneracy of the lowest energy band of \sto, affecting the Rashba effect on its surface \cite{nakamura2012}.

\section*{METHODS SUMMARY}
Preparation of \pary\ thin film was performed as follows.
In a $^\phi$46\,mm quartz tube, di-monochloro-\emph{para}-xylylene was sublimated at 150\,$^{\circ}\mathrm{C}$, the dimer flew into a high temperature (800\,$^{\circ}\mathrm{C}$) region \cite{williams}, where it was cleaved into two monomer units \cite{gorham}.
The reactive intermediates were then transported to a room temperature deposition chamber at the pressure of 2$\times$10$^{-4}$\,Pa.
Upon condensation on the surface of single-crystalline \sto, spontaneous polymerisation into \pary\ took place.
The \pary\ thin film deposited through this so-called Gorham method\cite{gorham} provides excellent conformal coating on the \sto\ surface.
The thin film of \TO\ was deposited directly on the top of the \pary\ by radio frequency (rf) sputtering at room temperature.
We used a \TO\ ceramic target in a flowing Ar gas of 25\,sccm and 5.5\,Pa.
The rf power was 2\,W/inch$^2$, and the sample-target distance was 140\,mm.
The thickness of the \TO\ layer was $\sim$200\,nm measured by a surface profiler (KLA Tencor).
On the top of \TO, a 150\,nm thick Au gate electrode with a 5\,nm Ti adhesion layer was deposited by electron-beam deposition.
The thickness of the \pary\ layer was estimated from the measured capacitance of a built-in capacitor (of area 800\,$\times$\,800\,$\mu$m$^{2}$), using the known values $\varepsilon$\,=\,3.2 for \pary, and $\varepsilon$\,=\,25 for \TO.

We used the Al metal for source and drain electrodes; Al gives ohmic contact to \sto.
For the other parts of the device fabrication process, we followed the same method and used the same materials as reported by Nakamura \etal\ \cite{nakamura1} 

For some samples, the thickness of the gate insulator was also measured by an S-4800 (Hitachi High-Tech) scanning electron microscope (SEM); the values were in good agreement with those estimated from the capacitance.
Since the \pary\ is much softer than \TO, when we tried to expose the cross sections for the SEM measurement, either cutting with a scalpel or simply cleaving, the bilayer at room temperature easily deformed (stretches and shrinks at the ends) the \pary\ layer, and the fibrils of \pary\ always extended out of the cross section.
We found that freezing the samples in liquid N$_2$ facilitates the cleaving.

Fig.~\ref{images}\,(b) shows the comparison of the cross section of a sample cleaved at room-temperature and one at liquid-N$_2$ temperature.
The \pary\ cross section in the latter case appears rather lumpy.
However, this was never observed in the cross sections of the samples cleaved at room-temperature; thus, it is likely that the uneven cross section is due to the steep rise of temperature with heavy bedewing in air after the cleavage in liquid-N$_2$.
It has been shown that a conventionally deposited planar \pary\ thin film \cite{gorham} does not possess nanostructured morphology \cite{damirel}.
\section*{Acknowledgements}
We are grateful to H. Shima, T. Yamada and T. Hasegawa for technical help. This work was supported by the Japan-Singapore Joint-Research Program, the Japan Society for the Promotion of Science (JSPS) and the National Research Foundation, Singapore through Competitive Research Programme (CRP Award No. NRF-CRP-4-2008-04). I.H.I. was partly supported by Grants-in-Aid for Scientific Research (category A, grant number 24244062).
\section*{Author Contributions}
I.H.I. and C.P. conceived and supervised the project. A.B.E. fabricated the devices and performed all the measurements. I.H.I. contributed to the experimental setup. P.S. and M.J.R. did the numerical simulation study. All the authors discussed the results and co-wrote the manuscript. Correspondence and requests for materials should be addressed to I.H.I.\,(\verb|i.inoue@aist.go.jp|) or C.P.\,(\verb|christos@ntu.edu.sg|).
\section*{Additional Information}
Supplementary Information accompanies this paper at
\href{http://www.nature.com/srep/2013/130424/srep01721/extref/srep01721-s1.pdf}{www.nature.com}.

%
%

%
\end{document}